\documentclass[10pt,a4paper]{article}

\usepackage{jheppub}

\usepackage{color}
\usepackage{colordvi}
\usepackage{changes}

\usepackage{epsfig,epsf}
\usepackage{amsmath}
\usepackage{amsthm}
\usepackage{amsfonts}
\usepackage{amssymb}
\usepackage{dsfont}
\usepackage{epstopdf}
\usepackage{multirow}
\usepackage{braket}
\usepackage{tabularx}

\usepackage{rotating}

\usepackage{marvosym}

\usepackage{todonotes}

\usepackage{subcaption}
\usepackage{array}   
\newcolumntype{C}{>{$}c<{$}}

\usepackage{slashed}

\usepackage[active]{srcltx}

\newcommand{\s}[1]{\slashed{#1}}

\newcommand{\w}[1]{\widetilde{#1}}
\newcommand{\f}[1]{\mathcal{#1}}

\newcommand{\tr}{\text{tr}\hspace{.05cm}}

\newcommand{\sbar}[1]{\slashed{\bar{#1}}}

\def\II{\hbox{{1}\kern-.25em\hbox{l}}}

\def\II{\hbox{{1}\kern-.25em\hbox{l}}}

\title{
Resummation of threshold logarithms in deeply-virtual Compton scattering}

\author[a]{J. Schoenleber}
\affiliation[a]{
   Institut f\"ur Theoretische Physik, Universit\"at
   Regensburg \\ D-93040 Regensburg, Germany}

\emailAdd{jakob.schoenleber@ur.de}

\abstract{
I derive an all-order resummation formula for the logarithmically enhanced contributions proportional to $\frac{\alpha_s^n}{x \pm \xi} \log (\frac{\xi \pm x}{2\xi})^k$ in the quark coefficient function of deeply-virtual-Compton scattering and the pion-photon transition form factor in momentum space. The resummation is performed at the next-to-next-to-leading logarithmic accuracy. The key observation is that the quark coefficient function itself factorizes in the $x \rightarrow \pm \xi$ limit, which allows for a resummation using renormalization group equations. A preliminary numerical analysis suggests that the corrections due to resummation for the quark contribution might be small.
}


\begin{document}
\maketitle

\section{Introduction}\label{sec:introduction}

Three-dimensional ``tomographic'' imaging of the proton is stated as a major scientific goal of the planned Electron-Ion collider (EIC) \cite{AbdulKhalek:2022hcn}. An important part of this project are the generalized parton distributions (GPDs), which encode information about the transverse spatial probability distribution of a given parton, carrying a given momentum fraction of the proton. The most prominent process giving access to GPDs is deeply virtual Compton scattering (DVCS)
\begin{align}
\gamma^*(q) + N(\text p) \rightarrow \gamma(q') + N(\text p'),
\end{align} 
where a hard virtual photon collides with a proton, which remains intact, and additionally, an outgoing photon is measured in the final state. We denote $Q^2 = -q^2, ~ t = (\text p-\text p')^2, ~ m^2 = \text p^2 = \text p'^2$.

In this work we consider the quark contribution to the Compton form factor $\f H_q,~q = u,d,s$. It is related to the quark GPD $H_q$, by the usual collinear factorization formula \cite{Radyushkin:1997ki, Collins:1998be, Ji:1998xh}
\begin{align}
\f H_q(\xi, Q, t) = \int_{-1}^1 \frac{dx}{\xi} C(x/\xi, Q, \mu) H_q(x,\xi,t, \mu).
\label{eq: DVCS factorization intro}
\end{align}
The coefficient function (CF) $C(x/\xi,Q,\mu)$ has been calculated recently to two-loop accuracy using two different approaches: based on conformal symmetry in \cite{Braun:2020yib}, and later by diagrammatic calculation in \cite{Braun:2022bpn}. The integration region of the convolution integral in this factorization formula includes the simple poles of $C$ at $x = \pm \xi$. These singularities are poles (with protruding branch cuts) in the partonic Mandelstam variables $\hat s = -\frac{1}{2} (1- x/\xi) Q^2$ and $\hat u = -\frac{1}{2} (1+ x/\xi) Q^2$ corresponding to the hard parton-photon process. Beyond the simple poles $C$ exhibits logarithmic divergences when $x \rightarrow \pm \xi$. The physical origin of this logarithmic enhancement is that the ratio of scales, $Q^2/\hat s$ for $x \rightarrow \xi$ and $Q^2/\hat u$ for $x \rightarrow -\xi$, becomes large when $\hat s$ or $\hat u$ become small.

The poles of $C$ in eq. (\ref{eq: DVCS factorization intro}) are formally avoided by deforming the integration contour into the complex plane according to the $\xi \rightarrow \xi - i0$ prescription \cite{Collins:1998be, Ji:1998xh}. This fails however at first sight, since the GPD $H_q$ has a discontinuous derivative at the points $x = \pm \xi$. Thus $H_q$ should be written as a sum of functions that are either analytic or have a simple zero at $x = \pm \xi$. For the analytic terms we can perform the contour deformation away from $x = \pm \xi$. However for the other terms the pole is at the endpoint of the integration contour.
This issue seemingly invalidates the use of the collinear approximation used to derive the factorization theorem. More precisely, if $|x\pm \xi|$ is comparable or smaller than $\frac{-t}{Q^2}, \frac{m^2}{Q^2}, \frac{|p_{\perp}^2|}{Q^2}$, where $p_{\perp}$ is the transverse momentum of the parton, i.e. when one of the momenta of the parton legs connecting the hard and collinear subgraph is in the soft (or Glauber) region, the collinear approximation fails. But, as was argued by Collins and Freund \cite{Collins:1998be}, the corresponding error made by applying the collinear approximation is power suppressed, so the leading twist factorization theorem remains valid. For these reasons, large logarithms of the form $\log(x\pm\xi)$ are formally avoided in eq. (\ref{eq: DVCS factorization intro}), so a resummation is not necessary to get good predictions. However, if one desires high precision, say at the low single digit percent level, it is necessary to check whether contributions from threshold resummation are sizable. 

An earlier attempt \cite{Altinoluk:2012nt} to resum threshold logarithms $\log (x \pm \xi)$ in $C$ claimed that the logarithms resum to a hyperbolic cosine.
The explicit calculation \cite{Braun:2020yib, Braun:2022bpn} disagreed with this statement and it was suggested that the leading (double) logarithms of $C$ in the $x \rightarrow \pm \xi$ limit  exponentiate. 

In this work we argue that exponentiation suggested in \cite{Braun:2020yib} holds to all orders and that in fact also higher order logarithms can be resummed. This procedure relies on the statement that the leading contribution to $C$ in the limit $x \rightarrow \pm \xi$ factorizes into a product of two functions that separate the large scale $Q^2 = -q^2$ and the center of mass energy of partonic process, i.e. $\hat s$ and $\hat u$. Given this factorization, renormalization group equations can be used to resum the logarithms. The result is a formula of the form
\begin{align}
C(x/\xi, Q, \mu = Q) &=\mp \frac{1}{2w_{\pm}}  \exp \Bigg \{ \frac{1}{2} \int_{Q^2w_{\pm} }^{Q^2} \frac{d\mu^2}{\mu^2} \Big [ - \Gamma_{\text{cusp}}(\alpha_s(\mu)) \log \Big ( \frac{w_{\pm} Q^2}{\mu^2} \Big ) + \bar \gamma_f(\alpha_s(\mu)) \Big ] \Bigg \} \nonumber
\\
&\qquad \times \bar h\Big(\alpha_s(Q)\Big) \bar f\Big(\alpha_s(\sqrt{w_{\pm}} Q) \Big ) + O(w_{\pm}^0 \times \text{logs}),
\label{eq: Cq final intro}
\end{align}
where $w_{\pm} = \frac{1}{2} (1 \pm x/\xi)$, $\Gamma_{\text{cusp}}$ is the well-known cusp anomalous dimension and $\bar f$ and $\bar h$ are functions that can be expanded in $\alpha_s$ with constant coefficients. We will show that this ansatz is in agreement with explicit calculation in \cite{Braun:2020yib, Braun:2022bpn} to $\alpha_s^2$ accuracy and the comparison allows one to extract the leading $\alpha_s^2$ contribution to $\bar \gamma_f$. The available ingredients allow for resummation up to the next-to-next-to-leading logarithmic (NNLL) accuracy, which is defined in Table \ref{tab: LL}. The corresponding expressions for $\bar h, \bar f$ and $\bar \gamma_f$ are collected in Appendix \ref{sec: f and h expressions}.

We argue further that the same formula applies to the axial vector case of DVCS or equivalently, to the CF of the pion-photon transition form factor. 

The threshold resummation in DVCS is closely related to usual cases of threshold resummation. In particular this applies to treatments in momentum space, such as the analysis of the endpoint $x_B\rightarrow 1$ region of DIS in \cite{Becher:2006mr}, where $x_B$ is conventional Bjorken variable $x_B = \frac{Q^2}{2\text p \cdot q}$, which is usually defined in the same way for DVCS, since these processes have the same initial state configuration. The crucial difference is that in the present treatment the kinematic limit where the CF factorizes is with respect to the partonic variable $x$, which is not fixed by external kinematics. In contrast, in the DIS case, the integration region is, by the external kinematical condition $x_B \approx 1$, restricted to a region where the parton momentum fraction is close to $1$. Thus, the hard scattering kernel factorizes when an external kinematical variable approaches some limit, contrary to DVCS, where the limit is with respect to a partonic variable. In this sense the present case is more similar to the well-known case of threshold resummation of Drell-Yan-like inclusive processes, see e.g. \cite{Beneke:2011mq, Sterman:2013nya}, where the hard scattering kernel factorizes when the partonic center of mass energy is close to some large final state mass scale. Consequently one encounters various difficulties related to the integration over a running scale in the resummed hard scattering kernel, when evaluating the convolution with the non-perturbative function. The same issue occurs in the present case, but if one has an analytic formula for the GPD, one can deform the contour of the $x$-integration into the complex plane, avoiding the Landau pole.

The presentation is organized as follows.
In Section \ref{sec:preliminaries} we introduce basic definitions. In Section \ref{sec: factorization} we derive a factorization theorem for the CF in eq. (\ref{eq: CF factorization}). In Section \ref{sec: oneloop}, we illustrate the general statement on the example of the one-loop calculation and confirm the corresponding prediction for the two highest power of logarithms at two-loops. In Section \ref{sec: axial} we briefly discuss the application to the axial case or equivalently the pion-photon transition form factor with the result that the same resummation formula applies. Finally, in Section \ref{sec: two-loop resum} we discuss how the NNLL resummation can be done and present the results of a preliminary numerical study.

\section{Preliminaries}\label{sec:preliminaries}

\begin{figure}
\centering
\includegraphics[scale=.53]{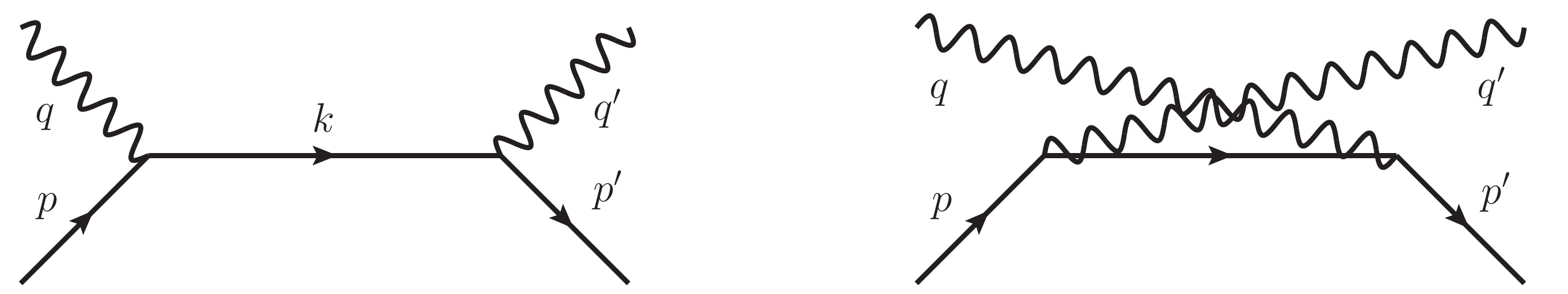}
\caption{Tree level diagrams for the DVCS CF. Left: $s$-channel, Right: $u$-channel}
\label{fig: treelevel}
\end{figure}
Throughout this work, we will treat mainly the vector case for DVCS, however the treatment for the axial-vector case is analogous and this is commented on in Section \ref{sec: axial}.
We consider the vector quark contribution to the Compton form factor $\f H_q$, which factorizes into the CF and quark GPD $H_q$ \cite{Radyushkin:1997ki, Collins:1998be, Ji:1998xh}
\begin{align}
\f H_{q}(\xi, Q, t) = \int_{-1}^1 \frac{dx}{\xi} C(x/\xi, Q, \mu) H_q(x,\xi,t, \mu).
\label{eq: DVCS factorization}
\end{align} 
The skewness parameter $\xi$ to leading twist accuracy can be chosen as \cite{Belitsky:2005qn}
\begin{align}
\xi = \frac{x_B}{2-x_B} + O(-t/Q^2),
\end{align}
where $x_B = \frac{Q^2}{2\text p \cdot q}$ is the Bjorken variable.

The tree-level diagrams contributing to $C$ are shown in Figure \ref{fig: treelevel}.
It follows from crossing symmetry that $C$ is anti-symmetric in the $x/\xi$ variable $C(x/\xi) = - C(-x/\xi)$. Indeed, each diagram has symmetric partner obtained by crossing the photon legs, i.e. the $s$- and $u$-channel. This crossing corresponds to the replacement $x/\xi \rightarrow - x/\xi$ and, in the vector case, to a relative minus sign. Hence it is enough to consider only one of the two channels and anti-symmetrize everything in the end. We will choose the $s$-channel, which is shown at tree level on the left in Figure \ref{fig: treelevel}.

It is convenient to introduce the variable $z = \frac{1}{2} (1 - x/\xi)$. In terms of this variable the crossing symmetry becomes $C(z) = - C(1-z)$. We can view $C(z)$ as a function of complex $z$ variable. It is analytic on $z \in \mathbb C\backslash((-\infty,0] \cup [1,\infty))$ and has simple poles at $z = 0$ and $z = 1$ with protruding branch cuts on the real axis on $z \not\in [0,1]$. 

From the conventional frame, where the in- and out-going parton momentum are given by $p = (x + \xi) P$ and $p' = (x-\xi) P$, where $P = \frac{\text p+ \text p'}{2}$, we make a Lorentz-boost $P^+ \rightarrow P^+/2\xi$. At leading twist kinematics we can neglect $t$ and the nucleon mass $m^2 = \text p^2 = \text p'^2$. Then $P = P^+ \bar n$ and
\begin{align}
p = (1-z) P^+ \bar n, \qquad 
p' = -z P^+ \bar n, \qquad 
q' = \frac{Q^2}{2P^+} n, \qquad 
q = - P^+ \bar n + \frac{Q^2}{2P^+} n ,
\end{align}
where we use two light-like vectors with $n \cdot \bar n = 1$ and $v^+ \equiv n \cdot v, ~ v^- \equiv \bar n \cdot v,~ v_{\perp} \equiv v - v^+ \bar n - v^- n$ for any four-vector $v^{\mu}$.\\
As a consequence of the crossing symmetry, the coefficients of the simple poles at $z \rightarrow 0$ and $z \rightarrow 1$, or equivalently $x \rightarrow \xi$ and $x \rightarrow -\xi$, are the same up to a minus sign. As mentioned before, for the rest of this work we will consider only the $s$-channel diagrams, which contain the leading singularities in the limit $z \rightarrow 0$.

We further introduce the momentum
\begin{align*}
k \equiv p + q = p' + q' = - z P^+ \bar n + \frac{Q^2}{2P^+} n,
\end{align*}
such that $k^2 = \hat s = - z Q^2$ is the invariant mass of the partons in the intermediate state. For $z \rightarrow 0$ we are considering the region where $k$ is close to being $n$-collinear, i.e. close to $q'$. The discussion for $z \rightarrow 0$ or $x \rightarrow -\xi$ is completely analogous and is obtained by the replacement $z \rightarrow 1-z$ or $\hat s \leftrightarrow \hat u$.
Throughout this work we are using Feynman gauge.

\begin{figure}
\centering
\includegraphics[scale=.38]{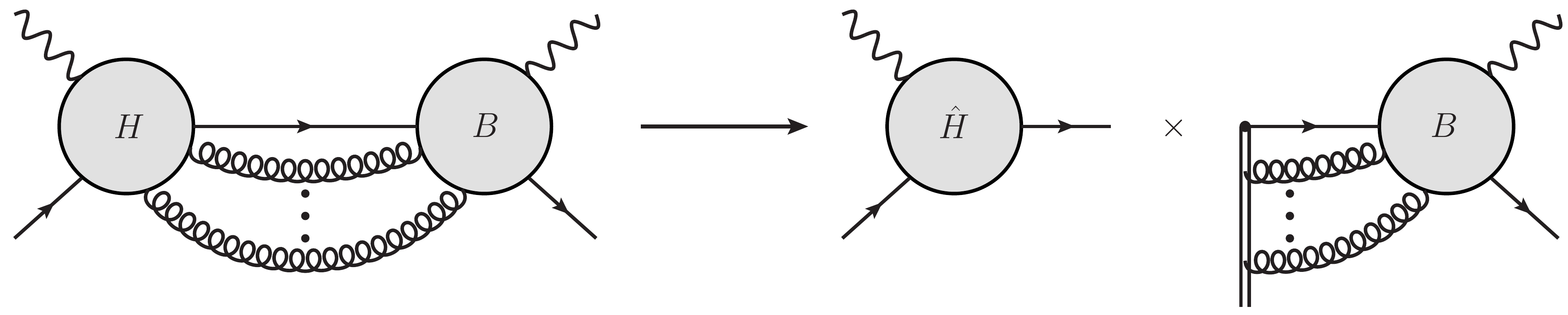}
\caption{Graphical representation of the factorization theorem in eq. (\ref{eq: CF factorization}). Left: Reduced graph for the leading region in the $z\rightarrow 0$ limit.  Right: Graphical representation of the left-hand-side of eq. (\ref{eq: CF factorization}). Shown is the product of the hard function $h$, defined in eq. (\ref{eq: h}), on the left and the $n$-collinear function $f$, defined in eq. (\ref{eq: f}), on the right. The double lines denotes the Wilson line $W_n$. Note that there are also diagrams with crossed gluons connecting to the Wilson line and a sum over hard subgraphs is implied.}
\label{fig: reduced dias}
\end{figure}

\section{Factorization of the coefficient function}
\label{sec: factorization}
We argue that in the $z \rightarrow 0$ limit the CF can be written in a factorized form
\begin{align}
C(Q^2, -k^2 ,\mu^2) = - \frac{Q^2}{2k^2}\Big [ h(Q^2, \mu^2) f(-k^2, \mu^2) + O(k^2/Q^2) \Big ].
\label{eq: CF factorization}
\end{align}
We will denote by $\f C$ twice the coefficient of the $\frac{1}{z}$ pole of $C$, i.e. $C = \frac{1}{2z} \f C + O(z^0)$. Then the factorization formula takes the form $\f C = hf$.
The bare quark vector CF of DVCS can be written as
\begin{align}
C^{\text{bare}}(Q^2, -k^2, \mu^2 ) &= -\frac{i g_{\perp}^{\mu \nu}}{4 (d-2)}  \int d^dx_1~ e^{iq'x_1}  \lim_{p \rightarrow (1-z)P} \lim_{p' \rightarrow - zP} \int d^dx_2~ e^{ip'x_2} \int d^dx_3 e^{-ip x_3} \nonumber 
\\
&\qquad \times \frac{1}{N_c} \tr \Big [ \s P  (-i \s p')\bra 0 T \{ \psi(x_2) j_{\nu}(x_1) j_{\mu}(0)  \bar \psi(x_3) \} \ket 0_{\text{connected}} (-i \s p)  \Big ],
\label{eq: CF expr}
\end{align}
where $j^{\mu} = \bar \psi \gamma^{\mu} \psi$ and the trace goes over color and Dirac indices and $N_c = 3$ is the number of colors. The external parton lines are considered to be inside the GPD and should be removed when calculating $C$. The amputation procedure is made explicit by writing $\lim_{p \rightarrow (1-z)P} \lim_{p' \rightarrow - zP} (-i \s p') ... (-i\s p)$. Note that the corrections on external legs are irrelevant, since the parton legs are on-shell. 

The renormalized CF, as is appears in eq. (\ref{eq: DVCS factorization}), is obtained from eq. (\ref{eq: CF expr}) by a subtraction of divergences corresponding to the renormalization of the GPD. Schematically $C = C^{\text{bare}} \otimes Z$, where $\otimes$ denotes convolution.\\ 
We start by appealing to the well-known Libby-Sterman analysis \cite{Libby:1978bx}, of which a review can be found in \cite{Collins:2011zzd}. We must first identify the leading regions and associated reduced graphs corresponding pinch-singular-surfaces, which are determined by the Landau criterion. In such reduced graphs one may have subgraphs with its lines being formally considered of the types hard, collinear to $n$ or $\bar n$, or soft. The major statement on which our treatment relies is that the $\bar n$-collinear and soft regions do not contribute. As is shown later, they are eliminated by setting the external quark momenta on-shell (then the contribution from those regions gives scaleless integrals), as is necessary for calculating hard coefficient functions. In DVCS the separation of the hard scale $Q^2$ from the small scales $-t,m^2$ has been performed in the factorization formula (\ref{eq: DVCS factorization}). $C$ itself corresponds to the hard subgraph from this perspective, but in the limit in region $z \rightarrow 0$, which is a region of the convolution integral in (\ref{eq: DVCS factorization}), we find a further hierarchy of scales, namely $-k^2 \ll Q^2$.

The argument that we do not have to consider any $\bar n$-collinear or soft subgraphs goes as follows. Firstly, any $\bar n$-collinear subgraph can depend only on the invariants $p^2,p'^2, p\cdot p'$, which are all zero in the on-shell limit. To demonstrate this, consider the following loop integral with loop momentum in the $\bar n$-collinear region
\begin{align}
\int_{l \sim P }  \frac{d^dl}{(l^2)^{n_1} [(l + P)^2]^{n_2} [(l + q')]^{n_3}}
\sim \int  \frac{d^dl}{(l^2)^{n_1} [(l + P)^2]^{n_1}(2l \cdot q')^{n_3}}
\propto \int_0^{\infty} \frac{d\alpha}{\alpha^{n_1 + n_2 + 1 - d/2}}.
\end{align}
The $\alpha$ integral gives zero in dimensional regularization. It is easy to see the same situation occurs for any $\bar n$-collinear loop momentum integration after all other integrations have been performed.
Note that we do not have to consider a numerator since it can be written as a sum of the same factors that appear in the denominator, so we get merely a sum of integrals of the same form.
Thus any such non-tree-level subgraph gives a scaleless integral and hence vanishes in dimensional regularization. A possible soft subgraph might attach to the $n$-collinear subgraph $B$ which must be at the outgoing photon vertex (soft lines connecting the hard subgraph give a power-suppression). However, by the same arguments as for example in DIS or DVCS \cite{Collins:1998be} the soft subgraph is not present. Correspondingly, the soft region for a generic loop integral gives scaleless integrals by a similar argument as for the $\bar n$-collinear region.

We proceed by using the standard Libby-Sterman power-counting formulas \cite{Collins:2011zzd} to determine the leading regions. In our case, where only a collinear and hard subgraph are present, the contribution from a given region $R$ is proportional to $\Big(\frac{Q^2}{-k^2}\Big)^{p(R)}$, where
\begin{align}
p(R) &= 4-\#(\text{external lines})
 - \#(\text{lines }B\text{ to }H) + \#(\text{scalar pol. gluon lines }B\text{ to }H).
\label{eq: pc rules}
\end{align}
This implies that the leading regions correspond to the reduced diagram shown on the left in Figure \ref{fig: reduced dias}, where the arbitrary number of collinear gluons connecting the $H$ and $B$ subgraph are scalar-polarized, i.e. the $g^{\mu \nu}$ in the gluon propagator is replaced by $n^{\mu} \bar n^{\nu}$.

Let us write the amplitude of the generic graph $\Gamma$ on the left in Figure \ref{fig: reduced dias} schematically as follows
\begin{align}
\Gamma(H,B) &= g_{\perp, \mu \nu}\int \prod_{j=1}^N d^dl_j~\frac{1}{N_c}\tr \Big [ \s P B_{a_1 ... a_N} ^{\nu \nu_1 ... \nu_N}(p',k,l_1,...,l_N)  \frac{i(\s k + \sum_j \s l_j)}{(k + \sum_j l_j)^2 + i0} \nonumber
\\
&\hspace{5cm} \times \Big (\prod_j g_{\mu_j \nu_j} \Big )  H_{a_1 ... a_N}^{\mu \mu_1 ... \mu_N}(p,k,l_1,...,l_N)  \Big ],
\label{eq: Gamma(H,B)}
\end{align}
where we routed the $N$ loop momenta $l_j$ of the gluon lines as entering the hard subgraph and going back through the single fermion line connecting $H$ to $B$. Conventional notation would absorb the intermediate fermion line into the $B$ subgraph, but we choose to make it explicit here. To obtain the leading term we need to apply the region approximator $T_{R(H,B)}$, which, when applied to a given graph, corresponds to making the following replacements:
\begin{itemize}
\item In the $H$ subgraph, replace $l_j \rightarrow \hat l_j = l_j^- n$ and $z \rightarrow 0$, i.e. $k \rightarrow q',~ p \rightarrow P$,
\item On the fermion line between the $H$ and $B$ subgraph we insert the projector $\frac{1}{2} \sbar n \s n$,
\item Replace $g_{\mu_j \nu_j} \rightarrow \frac{\hat l_{j,\mu_j} \bar n_{\nu_j}}{\hat l_j \cdot \bar n + i0}$.
\end{itemize}
Note that the expression for $\Gamma(H,B)$ in eq. (\ref{eq: Gamma(H,B)}) has been written in such a way that it corresponds to a region $R(H,B)$ (of loop-momentum space) where the lines in the $B$ subgraph are $n$-collinear and the lines in the $H$ subgraph are hard. We obtain
\begin{align}
T_{R(H,B)} \Gamma(H,B) &= g_{\perp, \mu \nu}\int \prod_{j=1}^N d^dl_j~\frac{1}{N_c}\tr \Big [ \s P B_{a_1 ... a_N} ^{\nu \nu_1 ... \nu_N}(p',k,l_1,...,l_N)  \frac{i(\s k + \sum_j \s l_j)}{(k + \sum_j l_j)^2 + i0} \nonumber
\\
&\hspace{3cm} \times \frac{1}{2} \sbar n \s n \Big (\prod_j \frac{\hat l_{j,\mu_j} \bar n_{\nu_j}}{\hat l_j \cdot \bar n + i0} \Big )  H_{a_1 ... a_N}^{\mu \mu_1 ... \mu_N}(P,q',\hat l_1,...,\hat l_N)  \Big ].
\end{align}

The next step is to use graphical Ward identities by taking into account the sum of diagrams with all possible connections of the $N$ gluons to $H$, in order to show that the collinear gluons decouple from $H$. This is relatively simple in the case of QED and somewhat more involved in the case of QCD. However the arguments have become standard, so we do not repeat them here. Note that this step requires to sum over all hard subgraphs of the same order in $\alpha_s$. This results in the expression
\begin{align}
\sum_H T_{R(H,B)} \Gamma(H,B) &= g_{\perp, \mu \nu}  \int \prod_{j=1}^N d^dl_j~ \frac{1}{N_c}\tr \Bigg \{ \sbar n B_{a_1 ... a_N}^{\nu \nu_1 ... \nu_N}(p',k,l_1,...,l_N) \s n \frac{i Q^2}{2k^2} \nonumber
\\
&\hspace{1cm} \times \Bigg [\sum_{\substack{\text{permutations} \\ \text{of }\{1,...,N\}}} (-g)^N  \frac{\bar n_{\nu_1} t_{a_1} ... \bar n_{\nu_N} t_{a_N}}{(l_1^- + i0) ... (\sum_{j=1}^N l_j^- + i0)} \Bigg ] \sum_{\hat H} \hat H^{\mu}(P,q')\Bigg \},
\label{eq: expr with momspace Wline}
\end{align}
where the $t_a$ are color matrices. The remaining sum goes over all possible subgraphs $\hat H^{\mu}(p,k)$, of the given order in $\alpha_s$, without collinear gluon insertions.
The expression in the square brackets can be identified to be a momentum-space Wilson line. In position space and in terms of gluon fields it reads
\begin{align}
W_{n}(x) = \text P \exp \Bigg [ ig \int_{-\infty}^0 ds ~ A^-(x + s \bar n) \Bigg ].
\end{align}
Note that $\hat H$ is a function of momenta which have zero transverse components. Thus
\begin{align}
\s n \hat H^{\mu}(P,q') \sbar n = c_{\hat H}(Q^2) \s n \gamma_{\perp}^{\mu} \sbar n,
\end{align}
where $c_{\hat H}$ is the contribution to the hard function $h$. Finally we arrive at the factorized form of $\Gamma$ summed over all possible hard subgraphs. It reads
\begin{align}
\sum_H T_{R(H,B)} \Gamma(H,B) = - \frac{Q^2}{2k^2} \sum_{\hat H} c_{\hat H}(Q^2) c_B(-k^2),
\end{align}
where 
\begin{align}
c_B(-k^2) &= - i g_{\perp, \mu \nu} \int \prod_{j=1}^N d^dl_j~ \frac{1}{N_c}\tr \Bigg \{\s n\gamma^{\mu} \sbar n B_{a_1 ... a_N}^{\nu \nu_1 ... \nu_N}(p',k,l_1,...,l_N)
 [...] \Bigg \}.
\end{align}
is the contribution to the function, which is denoted by $f$ in eq. (\ref{eq: CF factorization}). 
The above discussion implies that the bare $f$ can be written as a correlation function 
\begin{align}
f^{\text{bare}}(-k^2) &= \frac{2k^2}{Q^2} \frac{i g_{\perp}^{\mu \nu}}{4 (d-2)} \int d^dx_1~ e^{iq'x_1} \lim_{p' \rightarrow - z P} \int d^dx_2~ e^{ip'x_2} \nonumber
\\
&\hspace{2cm} \times  \frac{1}{N_c}\tr \Big [ \s P \gamma_{\mu}  (-i \s p')\bra 0 T \{ \psi(x_2) j_{\nu}(x_1) \bar \psi(0) W_n(0)  \} \ket 0_{\text{connected}} \Big ].
\label{eq: f}
\end{align}
A diagrammatic representation of $f$ is shown in Figure \ref{fig: reduced dias}. Since $f$ describes momentum modes that are collinear to the outgoing photon it is appropriate to call $f$ the $n$-collinear function in this context. For threshold resummation in the endpoint region of DIS the analogue of $f$ is the jet function, which is matrix element of a single quark field and a Wilson line $\bar \psi W_n$. $f$ differs markedly from the jet function, since there appears an additional $n$-collinear electromagnetic current in the correlator. 

The hard function $h$, on the other hand, can be identified as the Sudakov form factor with on-shell massless external legs
\begin{align}
\bra {q'} \bar \psi(0) \gamma^{\mu} \psi(0) \ket P_{\text{connected, amputated}} = \gamma_{\perp}^{\mu} h^{\text{bare}}(Q^2).
\label{eq: h}
\end{align}
or equivalently, the hard matching coefficient of the Sudakov form factor in momentum space, denoted by $\w C_V$ in \cite{Becher:2006mr}.

We have shown the factorization of the sum of a set of subgraphs, for a given region.
When summing over all graphs $\Gamma$ and regions $R$ one has to take into account the resulting double counting. On the graphical level one can define a subtraction procedure defined recursively over from smaller to larger regions in the sense of set inclusion. We refer to the standard literature on collinear factorization proofs. In the case presented here it is assumed that the double-counting subtractions can alternatively be formulated by ``renormalizing'' the corresponding ``bare'' functions, which is a standard procedure when using factorization theorems.
In terms of bare quantities we have the factorization formula
\begin{align}
\f C^{\text{bare}}(Q^2,-k^2) = h^{\text{bare}}(Q^2) f^{\text{bare}}(-k^2),
\end{align}
where $C^{\text{bare}} = \frac{1}{2z}\f C^{\text{bare}} + O(z^0)$. Note that renormalization of the $z^{-1}$ coefficient of the CF becomes multiplicative, i.e.
\begin{align}
C = C^{\text{bare}} \otimes Z =\frac{1}{2z} \f C^{\text{bare}} \f Z + O(z^0).
\label{eq: curly C def}
\end{align}
Note that we can ignore mixing with the gluon CF, since the pure-singlet quark contribution that generates this mixing is suppressed by an additional power of $z$. This is because such contributions must correspond to reduced graphs that have an additional quark line connecting the hard and collinear subgraph, leading to a suppression according to eq. (\ref{eq: pc rules}).

On the other hand, $h$ and $f$ can also be renormalized multiplicatively, i.e. $h^{\text{bare}} = h Z_h^{-1}$ and $f^{\text{bare}} = f Z_f$. Note however that $Z_h \neq Z_f$. In fact
\begin{align}
\f C^{\text{bare}} = \f C\f Z^{-1} = h^{\text{bare}} f^{\text{bare}} = h Z_h^{-1} Z_f f.
\end{align}
The renormalized quantities are (by definition) finite, so we must have $\f Z Z_h^{-1} Z_f = 1$ in minimal subtraction schemes, i.e. the $Z$ factors cancel, leading to the factorization theorem in terms of the renormalized quantities $\f C = hf$, as stated in eq. (\ref{eq: CF factorization}). The $\mu$-independence of $\f C^{\text{bare}}$ implies
\begin{align}
0 &= \frac{d}{d\mu} \f C(Q^2,-k^2,\mu^2) \f Z^{-1}(Q^2, -k^2, \mu^2) \nonumber
\\
&= \frac{d}{d\mu}  h(Q^2, \mu^2) \f Z^{-1}(Q^2, -k^2, \mu^2) f(-k^2,\mu^2).
\end{align}
Thus we are free to set the scale $\mu = Q$ and obtain
\begin{align}
\f C(Q^2,-k^2,Q^2) = \frac{h(Q^2, Q^2)  f(-k^2, Q^2)}{2z}.
\label{eq: CF expr2}
\end{align}

We remark that a different approach to proving eq. (\ref{eq: CF factorization}) is to treat the complete DVCS amplitude instead of just the CF itself. For this one must consider the virtuality $-k^2$ as an intermediate scale between the hard scale $Q^2$ and the small scales $-t,m^2$, i.e. $-t,m^2 \ll -k^2 \ll Q^2$. A proof could proceed in close analogy to \cite{Becher:2006mr} using soft-collinear effective theory (SCET), where, in addition to usual hard, collinear and soft degrees of freedom, $k$ would be classified as ``semi-hard'' and $p'$ as ``soft-collinear''. In the approach of this work, we essentially considered only the separation of the semi-hard scale $-k^2$ from the hard scale $Q^2$, and used that the regions corresponding to the other momentum scalings give scaleless integrals for on-shell partons.

\section{One-loop calculation}
\label{sec: oneloop}
\begin{figure}
\centering
\includegraphics[scale=.32]{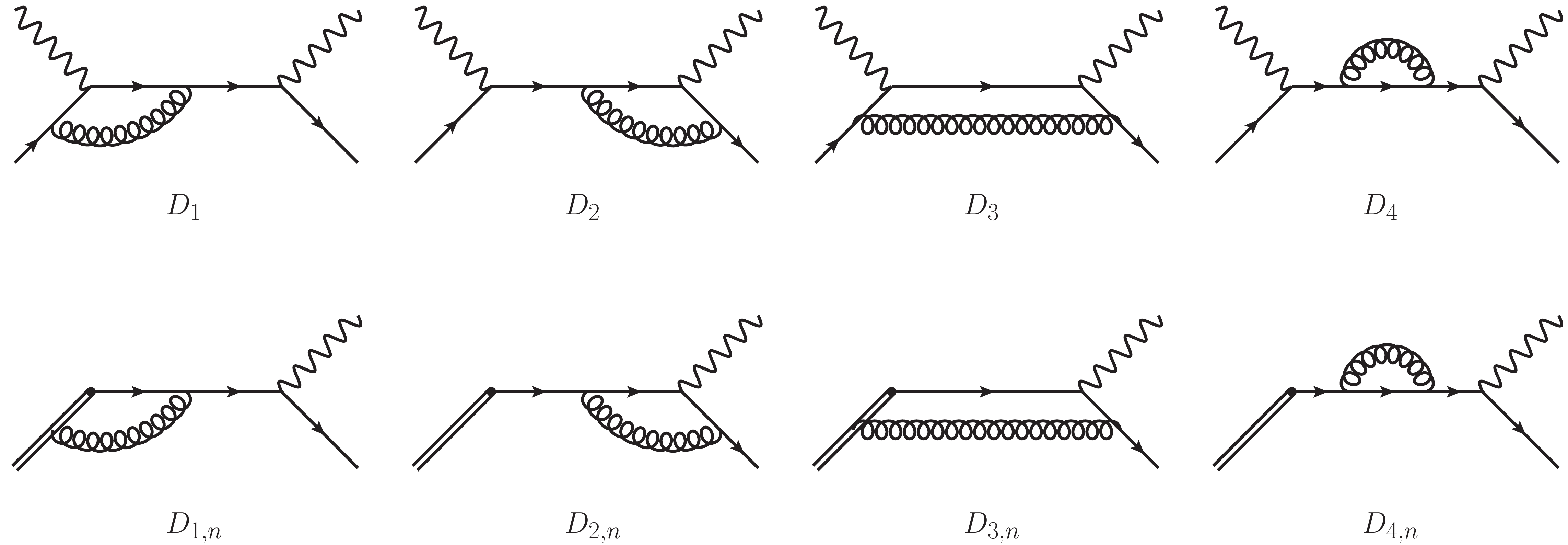}
\caption{Upper row: One-loop diagrams contributing to $C_q$, Lower row: One-loop diagrams contributing to $f$.}
\label{fig: oneloop dias}
\end{figure}
In this Section we verify the general statements above at one-loop accuracy. The diagrams are shown in the upper row of Figure \ref{fig: oneloop dias}. The expressions for the one-loop diagrams are well-known \cite{Ji:1998xh}. We quote the result when expanded in powers of $z$.
\begin{align}
D_1 &\simeq \frac{\alpha_s C_F}{4\pi} \Bigg [ - \frac{1}{\epsilon} \Big ( 1 + 2\log \frac{-k^2}{\mu^2} \Big ) - \log^2 \frac{Q^2}{\mu^2} + 3 \log \frac{Q^2}{\mu^2} + \log^2 \frac{-k^2}{\mu^2} - 2 \log \frac{-k^2}{\mu^2} - 4 \Bigg ],
\label{eq: D1res}
\\
D_2 &\simeq  \frac{\alpha_s C_F}{4\pi}  \Bigg [ - \frac{1}{\epsilon} + \log \frac{-k^2}{\mu^2} - 4 \Bigg ] ,
\label{eq: D2res}
\\
D_3 &\simeq 0,
\label{eq: D3res}
\\
D_4 &\simeq \frac{\alpha_s C_F}{4\pi} \Bigg [ - \frac{1}{\epsilon} + \log \frac{-k^2}{\mu^2} - 1 \Bigg ],
\label{eq: D4res}
\end{align}
where $\simeq$ means that we only keep terms of order $z^0$ times logarithms and we pulled out the tree-level amplitude $\frac{1}{2z} = - \frac{Q^2}{2k^2}$. 

The two relevant regions are $R_h$ and $R_n$, where the loop momentum is hard and $n$-collinear respectively. The corresponding region approximator was defined in Section \ref{sec: factorization}. We define $D_{j,h} = T_{R_h} D_j$ and $D_{j,n} = T_{R_n} D_j$. As $D_{2,n}$ and $D_{4,n}$ do not couple to the Wilson line, they are the same as $D_2$ and $D_4$, so the corresponding contribution from the hard region is zero. In particular they only depend on $-k^2$ and not on $Q^2$, as it should be. Diagram $D_3$ is suppressed by an additional power of $z$ and correspondingly $D_{3,n}$ vanishes since the Wilson line connects to the external leg, leading to a factor of $\bar n^2 = 0$.

The only one-loop diagram with a non-trivial matching is therefore $D_1$. We have
\begin{align}
D_{1,n} &= \frac{2k^2}{Q^2} \frac{ig_{\perp}^{\mu \nu}}{4(d-2)} (ig^2 C_F) \Big ( \frac{\mu^2 e^{\gamma_E}}{4\pi} \Big )^{\epsilon} \int \frac{d^dl}{(2\pi)^d} \frac{\text{tr}\Big [ \s P \gamma_{\nu} \s k \sbar n (\s k + \s l) \gamma_{\nu}\Big ]}{l^2 k^2 (l+k)^2 l^-} \nonumber
\\
&= - i g^2 C_F \Big ( \frac{\mu^2 e^{\gamma_E}}{4\pi} \Big )^{\epsilon}  \int \frac{d^dl}{(2\pi)^d} \frac{2P \cdot (k + l)}{l^2 (l+k)^2 (2P \cdot l)}  
\\
&= \frac{\alpha_s C_F}{4\pi} \Bigg [ \frac{2}{\epsilon^2} - \frac{1}{\epsilon} \Big ( 2 \log \frac{-k^2}{\mu^2} - 2 \Big ) + \log^2 \frac{-k^2}{\mu^2} - 2 \log \frac{-k^2}{\mu^2} + 4 - \frac{\pi^2}{6} \Bigg ].\nonumber
\end{align}
The contribution from the hard region, which is obtained by setting $z = 0$ in the integrand of $D_1$, can be found to be
\begin{align}
D_{1,h} &= \frac{\alpha_s C_F}{4\pi} \Bigg [ - \frac{2}{\epsilon^2} - \frac{3}{\epsilon} + \frac{2}{\epsilon} \log \frac{Q^2}{\mu^2} - \log^2 \frac{Q^2}{\mu^2} + 3 \log \frac{Q^2}{\mu^2} - 8 + \frac{\pi^2}{6} \Bigg ].
\end{align}
As expected, the sum $D_{1,n} + D_{1,h}$ reproduces the original result in eq. (\ref{eq: D1res}).

Eq. (\ref{eq: CF factorization}) allows us to resum the logarithms of $z$. As a simple illustration, I demonstrate how eq. (\ref{eq: CF expr2}) and the one-loop calculation can predict terms $\alpha_s^n \log^{2n-k}$ for $k = 0,1$ in $C$ to arbitrary orders. We have shown that
\begin{align}
f^{\text{bare}}(-k^2, \mu^2) &= 1 + \frac{\alpha_s C_F}{4\pi} \Bigg [ \frac{2}{\epsilon^2} - \frac{2}{\epsilon} \log \frac{-k^2}{\mu^2}  + \log^2 \frac{-k^2}{\mu^2} - 1 - \frac{\pi^2}{6} \Bigg ] + O(\alpha_s^2),
\end{align}
which implies that
\begin{align}
\frac{d}{d\log \mu} f(-k^2, \mu^2) = \Bigg [ - \frac{\alpha_s C_F}{\pi} \log \frac{-k^2}{\mu^2} + O(\alpha_s^2) \Bigg ] f(-k^2, \mu^2).
\end{align}
Solving this differential equation gives
\begin{align}
f(-k^2,Q^2) &= \exp \Bigg [ \frac{\alpha_s(Q) C_F}{4\pi} \log^2 z + \Big ( \frac{\alpha_s(Q)}{4\pi} \Big )^2 \Big ( - \frac{1}{3} \beta_0 C_F \log^3 z \Big ) \Bigg ] f(-k^2, -k^2) + ...,
\end{align}
where the ellipsis denote term that do not contribute to the two highest powers of logarithms.
Inserting this expression into eq. (\ref{eq: CF expr2}) gives
\begin{align}
C(z,\mu = Q) = \frac{1}{2z} \exp \Bigg [ \frac{\alpha_s(Q) C_F}{4\pi} \log^2 z + \Big ( \frac{\alpha_s(Q)}{4\pi} \Big )^2 \Big ( - \frac{1}{3} \beta_0 C_F \log^3 z \Big ) \Bigg ] + ...
\label{eq: CF resum1}
\end{align}
This result predicts the $\alpha_s^2 \log^4 z$ and $\alpha_s^2 \log^3z $ terms in the two-loop CF and the prediction indeed agrees with the explicit two-loop calculation of $C$ \cite{Braun:2020yib, Braun:2022bpn}. Of course we can resum more logarithms than presented in eq. (\ref{eq: CF resum1}). This is discussed in Section \ref{sec: two-loop resum}.

\section{Application to the axial-vector case and the pion-photon transition form factor}
\label{sec: axial}
The axial vector CF of DVCS in four dimensions is given by the same expression as eq. (\ref{eq: CF expr}) but with an additional $\gamma_5$ in the trace and projected onto the anti-symmetric part in $\mu \leftrightarrow \nu$. In $d = 4$
\begin{align}
\w C^{\text{bare}}(Q^2, -k^2, \mu^2 ) &= \frac{i \varepsilon_{\perp}^{\mu \nu}}{8}  \int d^dx_1~ e^{-iq'x_1}  \lim_{p \rightarrow (1-z)P} \lim_{p' \rightarrow - zP}\int d^4x_2~ e^{-ip'x_2} \int d^4x_3 e^{ip x_3}  \nonumber
\\
&\qquad \times \frac{1}{N_c} \tr \Big [ \gamma_5 \s P  (-i \s p)\bra 0 T \{ \psi(x_3) j_{\mu}(0) j_{\nu}(x_1) \bar \psi(x_2) \} \ket 0_{\text{connected}} (-i \s p')  \Big ],
\label{eq: axial Cq}
\end{align}
where $\varepsilon_{\perp}^{\mu \nu} = \varepsilon^{\mu \nu \rho \sigma} n_{\rho} \bar n_{\sigma}$.
Note that $\w C$ is the same coefficient function that appears for the pion transition form factor, with the only difference being the region of integration of the convolution being restricted to the range $0 < z < 1$. It is well-known that there is ambiguity in defining $\gamma_5$ in dimensional regularization. In the two-loop calculation of $\w C$ \cite{Braun:2021grd, Gao:2021iqq} Larin's scheme \cite{Larin:1993tq} was used and the result is then converted by a finite renormalization to the conventional $\overline{\text{MS}}$-scheme, which is defined by demanding that the evolution equation agrees with the vector case \cite{Braun:2021tzi}. Larin's scheme can be implemented by making the exchange
\begin{align}
\gamma_5 \s P \longrightarrow - \frac{i}{3!} \varepsilon_{\mu \nu_1 \nu_2 \nu_3} P^{\mu}  \gamma^{\nu_1} \gamma^{\nu_2} \gamma^{\nu_3}
\end{align} 
It is clear that this does not influence the arguments given in Section \ref{sec: factorization}, so the same factorization theorem applies
\begin{align}
\w C(Q^2, -k^2 ,\mu^2) = - \frac{Q^2}{2k^2} \Big [ \w h(Q^2, \mu^2) \w f(-k^2, \mu^2) + O(k^2/Q^2) \Big ]. 
\label{eq: CF fact axial}
\end{align}
In fact it can be found that for the result in \cite{Braun:2021grd, Gao:2021iqq} we get $f = \w f, ~ h = \w h$ up to two-loops (and possibly to higher orders). Hence the same formulas as in eq. (\ref{eq: CF resum1}) and (\ref{eq: Cq final}) apply to $\w C$.

\section{Resummation at NNLL accuracy}
\label{sec: two-loop resum}
By replacing $\log z = \log \frac{-k^2}{\mu^2} - \log \frac{Q^2}{\mu^2}$ in the two-loop result for $C$ \cite{Braun:2020yib, Braun:2022bpn} we can observe the separation of logarithms verifying eq. (\ref{eq: CF factorization}) at two loops. This also gives the two-loop expressions for $f$ and $h$ up to constants. The one-loop and two-loop expressions are collected in Appendix \ref{sec: f and h expressions}. 

In order to carry out the resummation we need the anomalous dimensions.
\begin{align}
\frac{df}{d\log \mu} = \gamma_f f, \qquad
\frac{dh}{d\log \mu} = \gamma_h h,
\qquad \frac{d\f C }{d\log \mu} = \gamma_{\f C} \f C.
\label{eq: f and h evo eq}
\end{align}
We have shown in Section \ref{sec: factorization} that $h$ is the hard matching coefficient of the Sudakov form factor.
It is well-known that its anomalous dimension has the all-order structure
\begin{align}
\gamma_h = \Gamma_{\text{cusp}} \log \frac{Q^2}{\mu^2} + \bar \gamma_h,
\end{align}
where the coefficient of the logarithm is the cusp anomalous dimension \cite{Korchemsky:1987wg}
\begin{align}
\Gamma_{\text{cusp}} = \frac{\alpha_s}{4\pi} 4 C_F + \Big ( \frac{\alpha_s}{4\pi} \Big )^2 \Big [ \frac{4}{3} (4- \pi^2) C_F C_A + \frac{20}{3} \beta_0 C_F \Big ] + O(\alpha_s^3).
\end{align}
Note that $\gamma_h + \gamma_f = \gamma_{\f C}$ can not depend on $\mu$, since the $\log \frac{Q^2}{\mu^2}$ logarithms in $\f C$ are single logarithms.
Hence the dependence on $\mu$ has to cancel in $\gamma_{\f C}$, which implies the all-order structure
\begin{align}
\gamma_f &= - \Gamma_{\text{cusp}} \log \frac{-k^2}{\mu^2} + \bar \gamma_f,
\\
\gamma_{\f C} &= - \Gamma_{\text{cusp}} \log z + \bar \gamma_{\f C},
\end{align}
where $\bar \gamma_{\f C} = \bar \gamma_h + \bar \gamma_f$. The one- and two-loop expressions for $\bar \gamma_f, \bar \gamma_h$ are collected in Appendix \ref{sec: f and h expressions}.

Let us turn to the evolution equation for the $n$-collinear function $f$, eq. (\ref{eq: f and h evo eq}), whose solution can be written as
\begin{align}
f(-k^2,Q^2) = U(z) f(-k^2,-k^2),
\label{eq: f sol}
\end{align} 
where
\begin{align}
U(z) = \exp \Bigg \{ \frac{1}{2} \int_{\log -k^2}^{\log Q^2} d\log \mu^2 \Big [ - \Gamma_{\text{cusp}}(\alpha_s(\mu)) \log \frac{-k^2}{\mu^2} + \bar \gamma_f(\alpha_s(\mu)) \Big ] \Bigg \}.
\label{eq: U factor}
\end{align}
Inserting eq. (\ref{eq: f sol}) into eq. (\ref{eq: CF expr2}) gives
\begin{align}
\f C(z, \mu = Q) = \frac{1}{2z} \bar h(\alpha_s(Q)) U(z) \bar f(\alpha_s(\sqrt{z} Q)),
\label{eq: Cq final}
\end{align}
where $\bar h(\alpha_s(\mu)) = h(\mu^2,\mu^2),~ \bar f(\alpha_s(\mu)) = f(\mu^2,\mu^2)$.
This implies that the result for the two-loop CF can be organized in the following way
\begin{align}
\f C^{(2)}(z, \mu = Q) &= \frac{1}{2!} \Big ( \frac{\Gamma_{\text{cusp}}^{(1)}}{4} \log^2 z \Big )^2 - \frac{\Gamma_{\text{cusp}}^{(1)}}{12} \beta_0 \log^3z + \frac{1}{4} \Big ( \Gamma_{\text{cusp}}^{(1)}\bar f^{(1)} + \Gamma_{\text{cusp}}^{(1)} \bar h^{(1)} + \Gamma_{\text{cusp}}^{(2)} \Big ) \log^2 z \nonumber
\\
&\hspace{2cm} - \Big ( \frac{\bar \gamma_f^{(2)}}{2} + \beta_0 \bar f^{(1)} \Big ) \log z + \text{const.}
\end{align}
For applications it can be convenient to rewrite the exponent in eq. (\ref{eq: U factor}) using
\begin{align}
\int_{\log \nu}^{\log \mu} d\log \mu' = \int_{\alpha_s(\nu)}^{\alpha_s(\mu)} \frac{d\alpha}{\beta(\alpha)}.
\end{align}
\begin{table}
\centering
\begin{tabular}{ |c|c|c|c|c|c|c| } 
\hline
RG-impr. PT & Log. approx. & $\sim \alpha_s^n \log^kz$ in $\log \f C$ & $\Gamma_{\text{cusp}}$ & $\bar \gamma_f$ & $\bar h, ~\bar f$ & $\beta$\\
\hline
- & LL & $n+1 \leq k \leq 2n$ & 1-loop & - & -& 1-loop\\
\hline
LO & NLL & $n \leq k \leq 2n$ & 2-loop & 1-loop & - & 2-loop\\ 
\hline
NLO &  NNLL & $n-1 \leq k \leq 2n$ & 3-loop & 2-loop & 1-loop & 3-loop\\
\hline
NNLO & NNNLL & $n-2 \leq k \leq 2n$ & 4-loop & 3-loop & 2-loop & 4-loop\\
\hline
\end{tabular}
\caption{Different approximation schemes. The logarithmic counting agrees with the one from \cite{Becher:2006mr}.}
\label{tab: LL}
\end{table}
This gives
\begin{align}
\log \f C &= 2S - A_{\bar\gamma_f} + \log \bar h + \log \bar f \nonumber
\\
&= \Big ( \frac{\alpha_s(Q)}{4\pi} \Big )^{-1} g_{\text{LL}}(z) + g_{\text{NLL}}(z) + \frac{\alpha_s(Q)}{4\pi} g_{\text{NNLL}}(z) + O(\alpha_s(Q)^2)
\label{eq: log C}
\end{align}
where
\begin{align}
S(z) &= \int_{\alpha_s( Q)}^{\alpha_s(\sqrt{z}Q)} d\alpha ~ \frac{\Gamma_{\text{cusp}}(\alpha)}{\beta(\alpha)} \int_{\alpha}^{\alpha_s(\sqrt{z}Q)} \frac{d\alpha'}{\beta(\alpha')},
\\
A_{\bar \gamma_f}(z) &= \int_{\alpha_s(Q)}^{\alpha_s(\sqrt{z}Q)} d\alpha~ \frac{\bar \gamma_f(\alpha)}{\beta(\alpha)}.
\end{align}
Explicit expressions for $g_{\text{LL}},g_{\text{NLL}},g_{\text{NNLL}}$ are collected in Appendix \ref{sec: g_j expressions}. The subscripts of the $g$-functions correspond to the logarithms that are resummed in it. The corresponding logarithmic counting scheme is defined in Table \ref{tab: LL}. Note that for the NNLL accuracy $\Gamma_{\text{cusp}}^{(3)}$ is required. Though it can not be obtained by methods used in the section, it is readily available in the literature. Hence, at this point in time, the highest accuracy that can be achieved is NNLL. Since $\Gamma_{\text{cusp}}^{(4)}$ and $\bar h^{(2)},\bar f^{(2)}$, see eqs. (\ref{eq: gammaf2}) and (\ref{eq: gammah2}), are also known, the only missing ingredient for NNNLL is $\bar \gamma_f^{(3)}$.

Let us consider how the result can be used in practice.
A naive way to implement the resummation corrections is to make the substitution
\begin{align}
C^{(\text{fixed order})}(z) \rightarrow C^{(\text{fixed order})}(z) &+ \frac{1}{2z} \Big ( \f C^{(\text{resummed})}(z) - 
\f C^{(\text{fixed order})}(z) \Big ) \nonumber
\\
&-  \frac{1}{2(1-z)} \Big ( \f C^{(\text{resummed})}(1-z) - 
\f C^{(\text{fixed order})}(1-z) \Big ),
\label{eq: prescription}
\end{align}
where $C^{(\text{fixed order})}$ is the CF to fixed order in perturbation theory and
\begin{align}
\f C^{(\text{resummed})}(z) = \exp \Bigg [ \Big ( \frac{\alpha_s(Q)}{4\pi} \Big )^{-1} g_{\text{LL}}(z) + g_{\text{NLL}}(z) + \frac{\alpha_s(Q)}{4\pi} g_{\text{NNLL}}(z) \Bigg ].
\end{align} 
The function $\f C^{(\text{fixed order})} $ is defined in such a way to subtract the double-counting of the terms of $\f C^{(\text{resummed})}$ that are already contained in $C^{(\text{fixed order})}$. It can be obtained by expanding $\f C^{(\text{resummed})}$ in $\alpha_s$ to the desired accuracy.

\begin{figure}
\centering
\includegraphics[scale=.3]{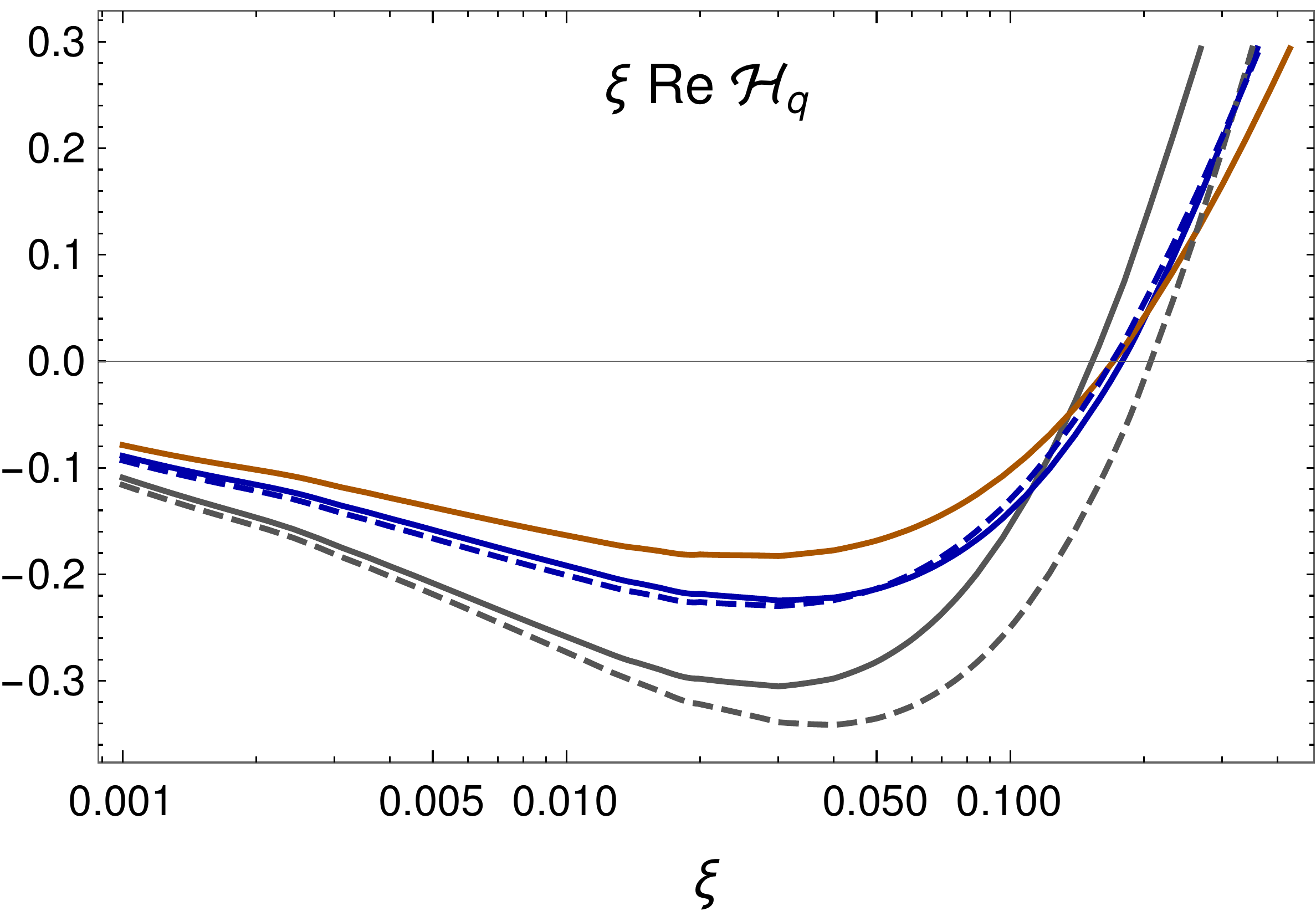}
\hspace{.1cm}
\includegraphics[scale=.3]{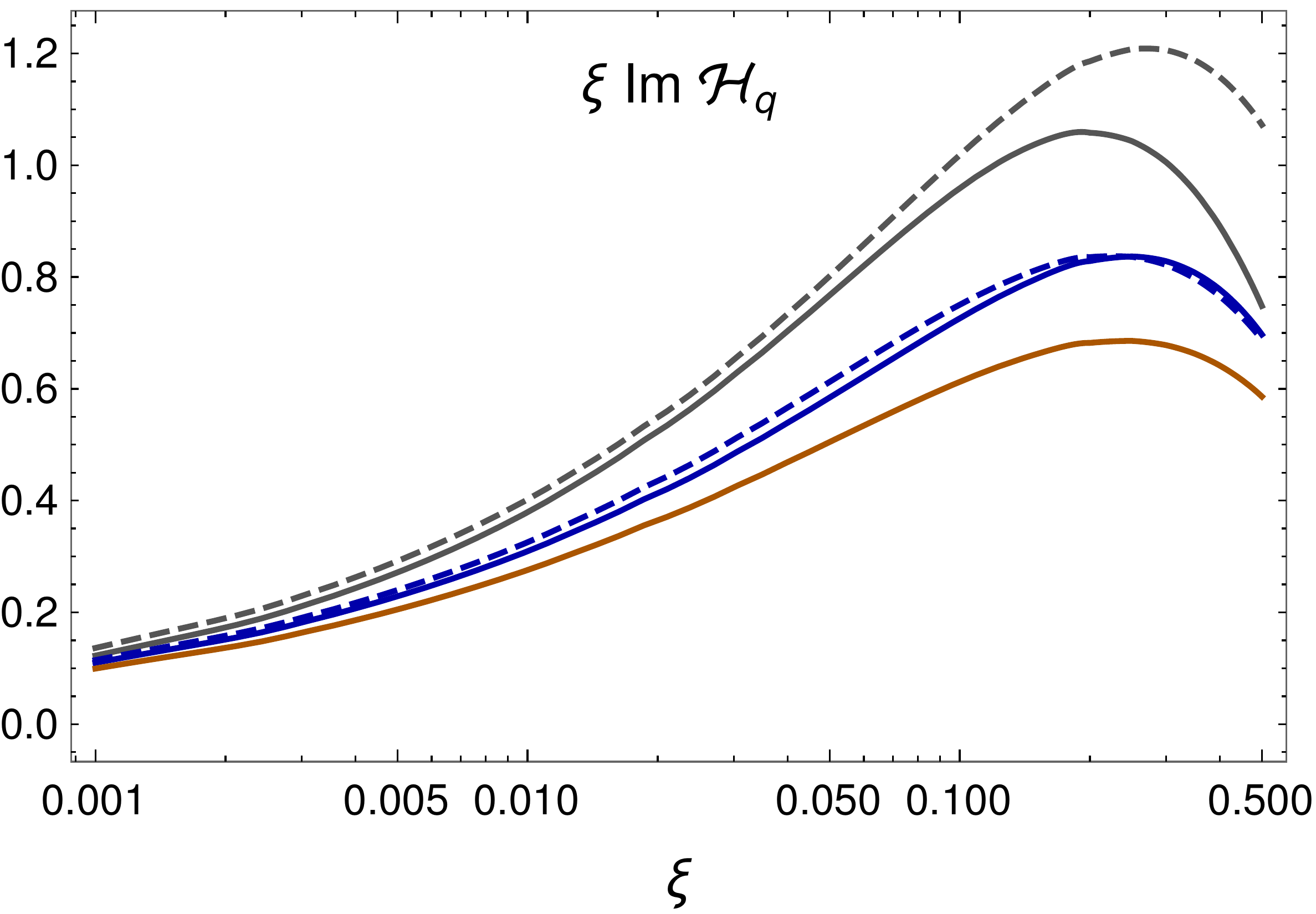}
\caption{Real and imaginary part of the quark CFF $\f H_q$, calculated using the model in eq. (3.331) of \cite{Belitsky:2005qn} for $n = 1/2$. The solid lines correspond to the CF at fixed order, gray being LO (fixed order LO and NLL resummation) and blue being NLO (fixed order NLO and NNLL resummation) while the dashed lines include the resummation, with the CF modified according to eq. (\ref{eq: prescription}). For reference, the plot of $\f H_{q}$ calculated with the fixed order NNLO non-singlet CF is shown in brown. The resummed NNLO result would require NNNLL resummation and not all the ingredients are known at this point. I have set $\mu = Q$, $n_f = 3$ and used $\frac{\alpha_s(Q)}{4\pi} = 0.025$.}
\label{fig: plots}
\end{figure}

The integration in eq. (\ref{eq: DVCS factorization}), when using the resummed CF, can be performed naively by deforming the contour into the complex plane, the same way as was done in \cite{Braun:2020yib, Braun:2022bpn}, when using the simple model quark GPD given in eq. (3.331) of \cite{Belitsky:2005qn} for $n = 1/2$, see Figure \ref{fig: plots}.

For this I used for the analytic form of the running coupling
\begin{align}
\alpha_s(\sqrt{z}Q) &= \frac{\alpha_s(Q)}{r} \Bigg \{ 1 - \frac{\alpha_s(Q)}{4\pi r} \frac{\beta_1}{\beta_0} \log r  \nonumber
\\
&\hspace{1cm}+ \Big ( \frac{\alpha_s(Q)}{4\pi r} \Big )^2 \Bigg [ \frac{\beta_1^2}{\beta_0^2} \Big ( \log^2 r - \log r - 1 + r \Big ) + \frac{\beta_2}{\beta_0} (1-r) \Bigg ] \Bigg \} + O(\alpha_s(Q)^4) 
\label{eq: alpha_s running}
\end{align}
where $r = 1 + \frac{\alpha_s(Q)}{4\pi} \beta_0 \log z$. This form of the running coupling organizes the perturbative expansion in terms of the leading log solution $\alpha_s^{(\text{LL})}(\sqrt{z} Q) = \frac{\alpha_s(Q)}{r}$. Thus the Landau pole is fixed at the point $r = 0$, or equivalently $z = e^{- \frac{4\pi}{\alpha_s(Q)\beta_0} }$.
It is clear that the points $z = 0$ and $z = 1$ never coincide with $r = 0$, so the contour can be deformed away from the Landau pole, making the numerical evaluation stable. Note that the $\xi \rightarrow \xi - i0$ prescription implies a natural direction of the contour deformation that should be performed in order to avoid the Landau pole.
 
As seen in Figure \ref{fig: plots} the correction due to the NNLL resummation appears to be small. This is because, as was mentioned in the introduction, the integration regions where the contour can not be deformed away from $z = 0$ or $z=1$, i.e. where threshold logarithms are large, are suppressed by powers of $z$. This is coherent with the discussion in \cite{Collins:1998be} which states that there is no leading power contribution from the region where $|x \pm \xi| \lesssim \frac{-t}{Q^2}, \frac{m^2}{Q^2}, \frac{|p_{\perp}^2|}{Q^2}$.

It should be mentioned that the estimates in Figure \ref{fig: plots} depend on the GPD model and the prescription of how to implement the resummation, in this case given by eq. (\ref{eq: prescription}). Furthermore there is dependence on the expression for the running coupling, in this case eq. (\ref{eq: alpha_s running}), and on how to deal with the Landau pole singularity. Therefore one can not conclude that the corrections from the threshold resummation are definitely small.
A more dedicated analysis should be performed in the future.

\section{Conclusion and outlook}
We have proposed a resummation formula  eq. (\ref{eq: Cq final}) for the quark CF in DVCS and the pion-photon transition form factor. The derivation relies on the factorization formula (\ref{eq: CF factorization}), which allows one to resum threshold logarithms using RG equations. The factorization theorem could be derived using standard arguments and it turns out to be a very simple case of factorization.
The general all-order arguments are backed by the two-loop calculation, where factorization was observed explicitly at two-loop accuracy. 

I mention that a SCET based proof, e.g. using the formalism in \cite{Becher:2006mr}  of (\ref{eq: CF factorization}) would be an interesting alternative to the approach in Section \ref{sec: factorization}. 

Although corrections due to the resummation to the quark Compton form factor appear to be small, see Figure \ref{fig: plots}, a dedicated analysis with a more realistic GPD model and a more careful investigation regarding how to treat $\alpha_s(\sqrt{z} Q)$ as an analytic function in $z$ will be performed in future work. A study of the impact of resummation for the pion photon transition form factor is interesting. Furthermore, employing the standard method of performing the resummation in moment space and comparing to the momentum space treatment considered in this work is also interesting.

Finally, a natural extension is to see whether a similar resummation can be done for the gluon contribution to $\f H$ and $\w{\f H}$. The corresponding two-loop expression for $\f H$ has been calculated in \cite{Braun:2022bpn}.

%
\begin{acknowledgments}
I thank Vladimir Braun and Werner Vogelsang for reading the manuscript and providing valuable insights. I also thank Valerio Bertone, Alexey Vladimirov, Yao Ji and Alexander Manashov for useful discussions. This work was supported by the Research Unit FOR2926 under grant 409651613.   
\end{acknowledgments}
%

\appendix

\section{Expressions for $f$ and $h$ functions}
\label{sec: f and h expressions}
We collect the results for the functions $f$ and $h$, defined in their bare form in eqs. (\ref{eq: f}) and (\ref{eq: h}).
We define $f = 1 + \frac{\alpha_s C_F}{4\pi} f^{(1)} + \Big ( \frac{\alpha_s C_F}{4\pi} \Big )^2 f^{(2)} + O(\alpha_s^3)$ and $h = 1 + \frac{\alpha_s C_F}{4\pi} h^{(1)} + \Big ( \frac{\alpha_s C_F}{4\pi} \Big )^2 h^{(2)} + O(\alpha_s^3)$. The one- and two-loop expressions are
\begin{align}
f^{(1)}(-k^2, \mu^2) &= C_F \log^2 \frac{-k^2}{\mu^2} + \bar f^{(1)},
\\
f^{(2)}(-k^2,\mu^2) &= \frac{1}{2} C_F^2 \log^4 \frac{-k^2}{\mu^2} -  \frac{1}{3}  \beta_0 C_F \log^3 \frac{-k^2}{\mu^2}  \nonumber
\\
&\qquad  + \Big [ - \Big ( 1 + \frac{\pi^2}{6} \Big ) C_F^2 +  \Big (\frac{4}{3}-\frac{\pi^2}{3} \Big ) C_F C_A + \frac{5}{3} \beta_0 C_F \Big ] \log^2 \frac{-k^2}{\mu^2} 
\\
&\qquad + \Big [ C_F^2 ( \pi^2 + 4 \zeta_3) - C_F C_A \Big (\frac{32}{9}-  14 \zeta_3 \Big )  - \frac{19}{9}  \beta_0 C_F \Big ] \log \frac{-k^2}{\mu^2} + \bar f^{(2)}, \nonumber
\\
h^{(1)}(Q^2, \mu^2) &= - C_F \log^2 \frac{Q^2}{\mu^2} + 3 C_F  \log \frac{Q^2}{\mu^2} + \bar h^{(1)},
\\
h^{(2)}(Q^2, \mu^2) &= \frac{1}{2} C_F^2 \log^4 \frac{Q^2}{\mu^2} + \Big ( - 3 C_F^2 + \frac{1}{3} \beta_0 C_F \Big ) \log^3 \frac{Q^2}{\mu^2}  \nonumber
\\
&\qquad + \Big [ \Big ( \frac{25}{2} - \frac{\pi^2}{6} \Big ) C_F^2 - \Big ( \frac{4}{3} - \frac{\pi^2}{6} \Big ) C_F C_A - \frac{19}{6} \beta_0 C_F  \Big ] \log^2 \frac{Q^2}{\mu^2} \nonumber
\\
&\qquad + \Big [ -\Big ( \frac{45}{2} + \frac{3\pi^2}{2} - 24 \zeta_3 \Big ) C_F^2 + \Big ( \frac{41}{9} - 26 \zeta_3 \Big )C_F C_A  
\\
&\hspace{3cm} + \Big ( \frac{209}{18} + \frac{\pi^2}{3} \Big ) \beta_0 C_F \Big ] \log \frac{Q^2}{\mu^2} + \bar h^{(2)}, \nonumber
\end{align}
where $\bar f^{(1)} = -C_F (1+\pi^2/6) $ and $\bar h^{(1)} = - C_F(8- \pi^2/6)$.
The on-shell massless Sudakov form factor $h$ is known, see e.g. eq. (50) and (51) in \cite{Becher:2006mr}, and I checked that the coefficients of the logarithms in $h^{(2)}$ obtained indirectly from $C^{(2)}$ agrees with those expressions. This is an important check of the formalism developed in Section \ref{sec: factorization}.

Although $\bar h^{(2)}$ and $\bar f^{(2)}$ do not contribute at NNLL, we give the results for completeness. Note that they can not be determined by the methods in Section \ref{sec: two-loop resum}. However $\bar f^{(2)}$ can be obtained from the result for $\bar h^{(2)}$, given in \cite{Becher:2006mr}, and the constant term of $\f C^{(2)}$, i.e. $\bar {\f C}^{(2)}= \bar h^{(2)} + \bar f^{(2)} + \bar h^{(1)} \bar f^{(1)}$, which can be obtained from the result in \cite{Braun:2020yib}. I find that
\begin{align}
\bar f^{(2)} &= C_F^2 \Big (\frac{3}{2} - \frac{\pi^2}{3} + \frac{119\pi^4}{360} - 39 \zeta_3 \Big ) + C_F C_A \Big ( \frac{95}{27} - \frac{4\pi^2}{9} - \frac{43\pi^4}{180} + 18 \zeta_3 \Big ) \nonumber
\\
&\hspace{1cm} + \beta_0 C_F \Big ( - \frac{7}{54} - \frac{5\pi^2}{36} - \frac{2}{3} \zeta_3 \Big ),
\\
\bar h^{(2)} &= C_F^2 \Big ( \frac{255}{8} + \frac{7\pi^2}{2} - \frac{83\pi^4}{360} - 30 \zeta_3 \Big ) + C_F C_A \Big ( - \frac{1037}{108}- \frac{7\pi^2}{9} + \frac{11\pi^4}{45} + 36 \zeta_3 \Big )  \nonumber
\\
&\hspace{1cm}+ \beta_0 C_F \Big ( - \frac{4085}{216} - \frac{23\pi^2}{36} - \frac{1}{3} \zeta_3 \Big ).
\end{align}
We present the expressions for the anomalous dimensions defined in eq. (\ref{eq: f and h evo eq})
\begin{align}
\gamma_f = \frac{\alpha_s}{4\pi} \gamma_f^{(1)} + \Big ( \frac{\alpha_s}{4\pi} \Big )^2 \gamma_f^{(2)} + O(\alpha_s^3),\qquad
\gamma_h = \frac{\alpha_s}{4\pi} \gamma_h^{(1)} + \Big ( \frac{\alpha_s}{4\pi} \Big )^2 \gamma_h^{(2)} + O(\alpha_s^3).
\end{align}
As explained in Section \ref{sec: two-loop resum} we have
\begin{align}
\gamma_f^{(j)} = -\Gamma_{\text{cusp}}^{(j)} \log \frac{-k^2}{\mu^2} + \bar \gamma_f^{(j)}, \qquad \gamma_h^{(j)} = \Gamma_{\text{cusp}}^{(j)} \log \frac{Q^2}{\mu^2} + \bar \gamma_h^{(j)}.
\end{align}
The expressions for the constant terms are $\bar \gamma_f^{(1)} = 0,~ \bar \gamma_h^{(1)} = -6C_F$ and
\begin{align}
\bar \gamma_f^{(2)} &= - 2 (\pi^2 + 4 \zeta_3) C_F^2 + \Big ( \frac{64}{9} - 28 \zeta_3 \Big ) C_F C_A +  \Big(\frac{56}{9} + \frac{1}{3} \pi^2 \Big ) \beta_0 C_F,
\label{eq: gammaf2}
\\
\bar \gamma_h^{(2)} &= - \Big ( 3 - 4\pi^2 + 48 \zeta_3 \Big ) C_F^2 - \Big (  \frac{82}{9} - 52 \zeta_3 \Big ) C_F C_A - \Big ( \frac{65}{9} + \pi^2 \Big ) \beta_0 C_F.
\label{eq: gammah2}
\end{align}

\section{Expressions for $g$ functions}
\label{sec: g_j expressions}
We give expression for the $g$ functions appearing in eq. (\ref{eq: log C}).
\begin{align}
g_{\text{LL}} &= \frac{\Gamma_{\text{cusp}}^{(1)}}{2\beta_0^2} (1-r+r\log r)
\\
g_{\text{NLL}} &= \frac{\Gamma_{\text{cusp}}^{(1)}}{4\beta_0^2} \frac{\beta_1}{\beta_0} \Big ( 2 - 2 r + 2 \log r + \log^2 r \Big ) - \frac{\Gamma_{\text{cusp}}^{(2)}}{2\beta_0^2} \Big (1 - r + \log r \Big ),
\\
g_{\text{NNLL}} &= \frac{1}{r} \Bigg \{ \bar f_1 + \bar h_1 r + \frac{\bar \gamma_f^{(2)}}{2\beta_0} (1-r) + \frac{\Gamma_{\text{cusp}}^{(1)}}{4\beta_0^2} \Bigg [ \frac{\beta_1^2}{\beta_0^2} \Big ( 1 - r + \log r \Big )^2 + \frac{\beta_2}{\beta_0} \Big ( 1- r^2 + 2 r \log r \Big ) \Bigg ] \nonumber
\\
&\qquad + \frac{\Gamma_{\text{cusp}}^{(2)}}{4\beta_0^2} \frac{\beta_1}{\beta_0}  \Big ( 3 - 4 r + r^2 + 2 \log r \Big ) + \frac{\Gamma_{\text{cusp}}^{(3)}}{4\beta_0^2} (1-r)^2 \Bigg \},
\end{align}
where $r = 1 + \frac{\alpha_s(Q)}{4\pi} \beta_0 \log z$.




\providecommand{\href}[2]{#2}\begingroup\raggedright\endgroup

\end{document}